\documentclass[aps,pre,twocolumn,showpacs,nofootinbib]{revtex4-1}

\usepackage{xcolor}
\definecolor{color1}{RGB}{0,150,0} % Color of the article title and sections
\definecolor{color2}{RGB}{0,0,150} % Color of the boxes behind the abstract and headings

\usepackage{amsmath,amssymb,amsfonts,graphicx}
\usepackage{hyperref}
\hypersetup{hidelinks,colorlinks,breaklinks=true,urlcolor=color2,citecolor=color1,linkcolor=color2,bookmarksopen=false,pdftitle={\@title},pdfauthor={\@author}}

\begin{document}

\title{Tearing of thin sheets: Cracks interacting through an elastic ridge}

\author{Fabian Brau}
\email{fabian.brau@ulb.ac.be}

\affiliation{Nonlinear Physical Chemistry Unit, Facult\'e des Sciences, Universit\'e libre de Bruxelles (ULB), CP231, 1050 Brussels, Belgium}

\begin{abstract} 
We study the interaction between two cracks propagating quasistatically during the tearing of a thin brittle sheet. We show that the cracks attract each other following a path described by a power law resulting from the competition between elastic and fracture energies. The power law exponent ($8/11$) is in close agreement with experiments. We also show that a second (asymptotic) regime, with an exponent of $9/8$, emerges for small distances between the two crack tips due to the finite transverse curvature of the elastic ridge joining them.
\end{abstract}


\date{\today}

\maketitle

\section{Introduction}
\label{sec:intro}

Cracks and fractures are very common phenomena occurring in various contexts~\cite{ande05,bouc10,bueh10,bouc14}. They are observed during the desiccation of films made of colloidal suspensions, like bentonite clay or cornstarch~\cite{alla95,bohn05a,bohn05b,goeh09,goeh10,laza11,boul13}, in sol-gel films~\cite{send03,mart14}, in broken objects like windows~\cite{astr97,kun06,vand13} or in sea ice~\cite{weis03,kors04} and ice floes collisions~\cite{vell07,vell08}. 

A material fractures when sufficient stress is applied at the level of its elementary constituents to break the bonds that hold them together. This process occurs mainly at the atomic scale near the crack tip, where the energy focuses, but also at much larger scale for particle rafts~\cite{vell06}. Nevertheless, macroscopic parameters, like work of fracture $\gamma$ or fracture toughness $K$, can be defined (and measured) to describe the progression of cracks when the material properties are uniform without necessarily resorting to microscopic analysis~\cite{ande05,lawn93,freu98}. The classical fracture theories, initially formulated by Griffith and Irwin~\cite{grif21,irwi57}, reliably describe the onset of crack motion but there is no general theory able to predict the path of a crack as it propagates. Understanding and predicting the propagation of a crack in a brittle material is a central challenge in fracture mechanics~\cite{freu98}. 

There are three ways of applying a force to enable a crack to propagate: in-plane tensile or shear loading (opening or sliding mode) and out-of-plane shear loading (tearing mode). Thin films offer an efficient setup to study the tearing mode with some practical interests since it is a natural mode to torn thin sheets~\cite{roma13,rome13}. Important insight about crack paths has been gained in this context thanks to the limitation of the crack motion to a two-dimensional manifold. For example, the crucial role of geometry was identified in some oscillatory fracture patterns obtained when a brittle elastic thin sheet is cut by a moving blunt object~\cite{roma03,ghat03,audo05,atki07,tall11}. It was also shown that a pair of cracks propagating and interacting in thin sheets subjected to in-plane tensile stress forms universal shapes~\cite{lang68,swai78,fend10}.

\begin{figure*}
\includegraphics[width=\textwidth]{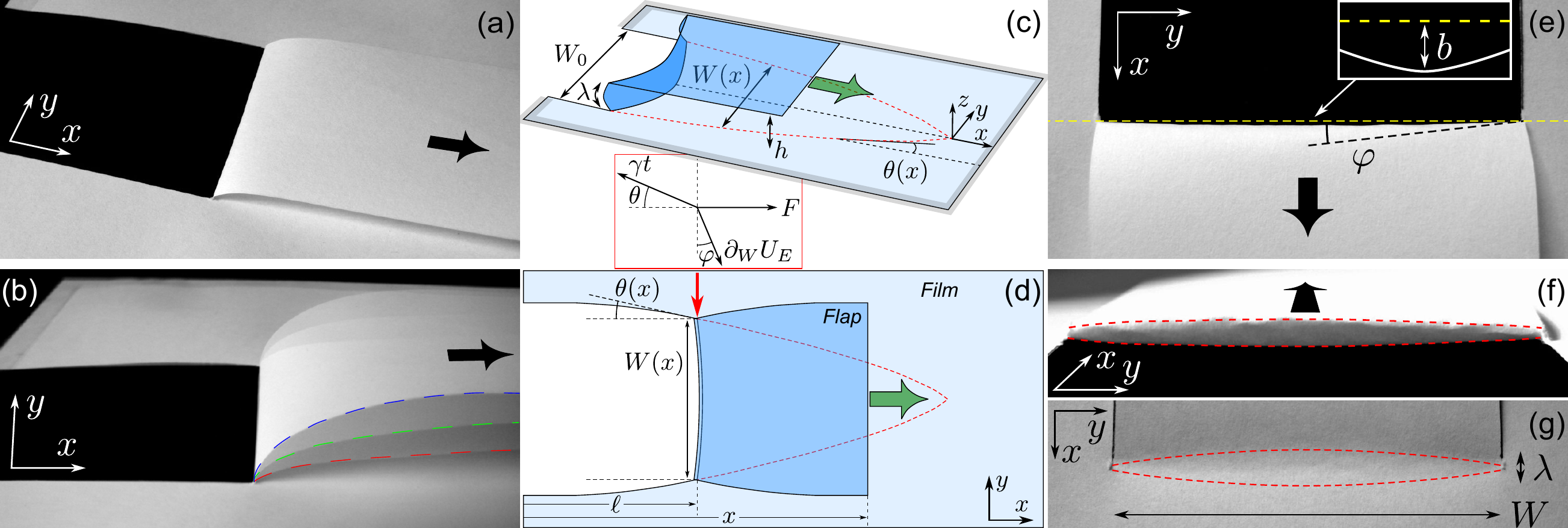}
\caption{(color online) Pictures and schematics of the tearing of a thin sheet. A sheet of paper is used to better illustrate the system. (a) Initial stage just before the crack propagation. The large arrow indicates the direction of the applied force. (b) Superimposition of three pictures showing the evolution of the ridge shape as the applied force, $F$, increases up to the onset of crack propagation. Colored dashed lines indicate the border of the flap for better visualization. (c-d) Schematics showing the variables needed to describe the system. The curved dashed lines indicate the path followed by the two cracks upon pulling the flap in the direction shown by the large arrow. The balance of forces at the crack tip is shown in the red rectangle. (e-g) Ridge morphology. (e) Upper view showing the transverse curvature of the ridge together with the angle $\varphi$ and the sag of the ridge $b$. (f) Longitudinal view along the $x$-axis showing the height $h$ of the ridge together with its two pinched edges. (g) Shape of the ridge which remains (slightly) visible once it is unfolded thanks to some plastic deformations in the paper sheet. $W=5$ cm in all pictures.}
\label{fig01}
\end{figure*}

In this work, we consider the tearing of a clamped thin brittle sheet (see Refs.~\cite{tall11,atki95} for ductile sheets) where two cracks interact during their propagation induced by the force applied on a rectangular flap peeled with a given peeling angle, see Fig.~\ref{fig01}. This system has already been studied when the sheet adheres to a flat substrate. In this case, the balance between fracture, adhesion and bending energies yields to converging linear crack paths~\cite{hamm08}. This system has been used to study mechanical properties of graphene~\cite{sen10} and to show that the curvature of the substrate modifies the crack paths, leading even to diverging trajectories~\cite{krug11}. Here we study the situation where adhesion is negligible. It has been shown experimentally that the crack paths are no longer linear and follow power laws with characteristic exponents: $3/4$ in the ``peeling" configuration with a peeling angle equals to $\pi$ and $2/3$ in the ``trousers" configuration~\cite{baya10,baya11}. Surprisingly, in contrast with results obtained for adhesive sheets, the theory developed in Ref.~~\cite{baya11} predicts that the crack paths are independent on the material properties and scale only with the sheet thickness.

We revisit this system in the peeling configuration by using the formalism developed in Ref.~\cite{hamm08} and by analyzing the elastic energy of the film essentially contained in the ridge joining the two cracks. We find that both elastic and fracture energies determine the crack paths.

\section{Setup and main equations}
\label{sec:setup}

Figure~\ref{fig01} shows pictures and schematics of the system under consideration. A thin film is clamped on a flat plate with narrow adhesive tapes along its borders. There is no significant adhesion between the film and the plate. Two parallel notches, separated by a distance $W_0$, are cut on one of its edges such that a rectangular flap is created. The flap is pulled with a peeling angle equals to $\pi$ at constant slow speed (in the range $0.05-1.5$ mm/s~\cite{baya11}) leading to a quasistatic crack propagation. The two crack tips move both forwards along the $x$-axis and inwards (towards $y=0$) until they eventually annihilate. A pointy flap is then detached from the film, see inset of Fig.~\ref{fig02}.

The pulling force $F$ applied to the flap deforms the fold joining the flap to the film such that, at the onset of crack motion, a small ridge focusing the elastic energy is formed, see Fig.~\ref{fig01}(b). The shape of this ridge is shown in Fig.~\ref{fig01}(e)-(g). It possesses two curvatures: one in the longitudinal direction joining the flap to the film and another one in the transverse direction along the $y$-axis with a maximal deviation from a straight line denoted $b$ (sag of the ridge), see Fig.~\ref{fig01}(e). This second curvature is due to the pulling force $F$ which applies along the entire width of the ridge whereas the resistive fracture force applies only at its edges where the crack tips are located. This leads also to the formation of two pinched edges, see Fig.~\ref{fig01}(f). Therefore, the ridge possesses the characteristics of a Lobkovsky-Witten ridge~\cite{witt93,lobk95,lobk96,lobk97,venk04} which appears generically between two points of high curvature in thin sheets~\cite{witt97}. Notice that for the tearing of adhesive sheets, the corresponding fold does not possess a transverse curvature because the adhesive force applies along its whole width and prevents any transverse bending. The elastic energy stored in the ridge can be released in two ways: by decreasing the longitudinal curvature of the ridge (advancing the crack in the pulling direction) or by simply reducing the width of the ridge (the cracks move inwards). The actual direction followed by the cracks is a combination of both effects.

The standard formalism we used to describe the system has been introduced in Ref.~\cite{hamm08} and subsequently used, for example, in Refs.~\cite{mart14,rome13,roma13}. It is briefly recalled here for self-containedness. The total energy of the system is
\begin{equation}
\label{energy}
U=U_{\rm{E}} + 2\gamma t s,
\end{equation}
where the first term is the elastic energy, which is essentially focused in the ridge, and the second one is the fracture energy for the two cracks. $t$ is the film thickness, $s$ is the crack length and $\gamma$ is the work of fracture of the film. The position of the crack tips is denoted $\ell$ and the position of the border of the flap where the pulling force $F$ is applied is denoted $x$, see Fig.~\ref{fig01}(d). The excess of length $2\ell-x = \lambda$ is the length of the ridge, see Fig.~\ref{fig01}(c),(g). As shown in the next section, the elastic energy of the ridge depends only on its width, $W$, and its length, $\lambda$: 
\begin{equation}
\label{ener-elas-gen}
U_{\rm{E}}=U_{\rm{E}}(\lambda=2\ell-x,W).
\end{equation}

In order to derive the relevant equations in a simple way, we first neglect the transverse curvature of the ridge ($\varphi=0$). The crack tip moves to a position that minimizes the total energy~\cite{ande05,lawn93,freu98}. For a displacement-controlled experiment, the requirement that the energy is minimal, ${\rm d} U/{\rm d} s=0$, together with Eq.~(\ref{energy}) yields the condition
\begin{equation}
\label{min-ener}
-2 \partial_W U_{\rm{E}}\sin \theta+ \partial_{\ell} U_{\rm{E}} \cos \theta +2\gamma t = 0,
\end{equation}
where ${\rm d} \ell/{\rm d} s = \cos \theta$ and ${\rm d} W/{\rm d} s = -2\sin \theta$ (by convention a positive $\theta$ corresponds to a decrease of $W$ as the crack advances). This equation is simply the balance of forces projected along the crack direction. In addition, the pulling force applied to the flap at position $x$ is given by the work theorem as $F = \partial_x U_{\rm{E}}$ for a quasistatic fracture propagation. Using Eq.~(\ref{ener-elas-gen}), we obtain the identities 
\begin{equation}
\label{work}
F = \partial_x U_{\rm{E}} = -\partial_{\lambda} U_{\rm{E}} = -\frac{1}{2} \partial_{\ell} U_{\rm{E}}.
\end{equation}
Combining Eq.~(\ref{min-ener}) with Eq.~(\ref{work}) leads to the following expression for the force
\begin{equation}
\label{force-gen}
F =\frac{\gamma t - \partial_W U_{\rm{E}} \sin \theta}{\cos \theta}.
\end{equation}
The fracture path is obtained by requiring that the tear follows the direction where the force is minimal for the advancement of the crack tips, $\partial_{\theta}F=0$. A differentiation of Eq.~(\ref{force-gen}) with respect to $\theta$ gives the direction followed by the cracks
\begin{equation}
\label{sintheta}
\sin \theta = \partial_W U_{\rm{E}}/(\gamma t).
\end{equation}
Substituting Eq.~(\ref{sintheta}) in Eq.~(\ref{force-gen}) gives
\begin{equation}
\label{force-gen2}
F = \gamma t \sqrt{1-\left[\partial_W U_{\rm{E}}/(\gamma t)\right]^2}=-\partial_{\lambda} U_{\rm{E}},
\end{equation}
where we also used Eq.~(\ref{work}).

Once the expression of the elastic energy $U_{\rm{E}}(\lambda,W)$ of the fold is known, Eq.~(\ref{force-gen2}) gives the expression of the ridge length $\lambda$ as a function of its width $W$ and the material constants ($\gamma$, $t$, Young modulus $E$ and Poisson ratio $\nu$). Substituting this expression of $\lambda$ into Eq.~(\ref{sintheta}) gives then the expression of $\theta$ as a function of $W$ and the material constants. Since $\theta$ is the local angle between the tangent to the crack path and the $x$-axis, the path is determined from the differential equation ${\rm d} W/{\rm d} \ell = -2\tan \theta(W)$ with the initial condition $W(0)=W_0$. However, it is more convenient to place the point where the two cracks meet at the origin of the coordinates and to consider the increase of the distance $W$ between the two cracks as a function of the distance to the origin (which we still denote $\ell$ for simplicity). This is achieved with the differential equation
\begin{equation}
\label{ode}
\frac{{\rm d} W}{{\rm d} \ell} = 2\tan \theta(W) \quad {\rm and} \quad W(0)=0.
\end{equation}

To obtain the relevant equations for a finite transverse curvature of the ridge, we notice that Eqs.~(\ref{sintheta}) and (\ref{force-gen2}) are equivalent to~\cite{hamm08} 
\begin{equation}
\label{eq-phi-nul}
F = \gamma t \cos \theta \quad {\rm and} \quad \partial_W U_{\rm{E}} = \gamma t \sin \theta,
\end{equation}
which correspond to the projections of the forces along the $x$ and $y$-axis as shown in Fig.~\ref{fig01}(c),(d) (when $\varphi=0$). Therefore, a finite transverse curvature of the ridge modifies Eqs.~(\ref{eq-phi-nul}) as follow
\begin{subequations}
\label{main-eq}
\begin{align}
\label{force}
F + \partial_W U_{\rm{E}} \sin \varphi &= \gamma t \cos \theta, \\
\label{theta}
\partial_W U_{\rm{E}} \cos \varphi &= \gamma t \sin \theta.
\end{align}
\end{subequations}
Notice however that, as shown below, the influence of the angle $\varphi$ is essentially negligible except in a small region, $W \ll W_{{\rm c}}$, near the tip of the detached flap where the two cracks meet. In the next section, we show that $\varphi$ depends only on $\lambda$ and $W$. Therefore, Eqs.~(\ref{force}) and (\ref{work}) give the expression of the ridge length $\lambda$ as a function of $W$ and $\theta$ (and the material constants). This expression of $\lambda$ is then used in Eq.~(\ref{theta}) to obtain the expression of $\theta$ as a function of $W$. The crack paths are finally determined by solving Eq.~(\ref{ode}).

\section{Elastic energy}
\label{sec:energy}

In order to compute explicitly the crack paths, we need to obtain the elastic energy of the system. As seen in Fig.~\ref{fig01}(b), the elastic energy focuses in a small folded region joining the flap to the film between the two crack tips as the applied force increases up to the onset of crack displacement. As mentioned above and seen in Fig.~\ref{fig01}(e)-(g), this folded region, containing essentially all the elastic energy, possesses the characteristics of a Lobkovsky-Witten ridge. We assume that such a ridge describes the elastic energy of our system. Notice that, if the stretching modulus, $Et$, is low or the fracture energy $\gamma$, is large, the flap could stretch significantly when it is pulled. Therefore, if $Et/\gamma \lesssim 1$, this additional stretching energy should be taken into account (see Ref.~\cite{sen10} for such an extension of the theory in the case of adhesive sheets). The experiments we consider are characterized by $Et/\gamma \gg 1$, and we thus assume that the elastic energy is mainly focused in the ridge.

The geometry and the elastic energy of the Lobkovsky-Witten are known~\cite{witt93,lobk95,lobk96,lobk97,venk04} and are recovered in the Appendix using a simple scaling approach:
\begin{subequations}
\label{ridge}
\begin{align}
\label{ridge-ener-temp}
& U_{\rm{E}} = C_{{\rm R}} B\, (W \alpha^7 /t)^{1/3}, \\
\label{ridge-length}
& \lambda \simeq h = C_{\lambda} (W^2 t/\alpha)^{1/3},
\end{align}
\end{subequations}
where $C_{{\rm R}} = {\cal R}(12(1-\nu^2))^{1/6}$ with ${\cal R} = 1.20\pm 0.04$~\cite{lobk96}, $B=Et^3/(12(1-\nu^2))$ is the bending modulus and $h$ is the height of the ridge in the $z$-direction and is proportional to its length $\lambda$ (see Fig.~\ref{fig01}(c)). The constant $C_{\lambda}$ is unknown and is considered as a free parameter of order~1. The parameter $\alpha$ is the dihedral angle of the ridge (see Fig.~\ref{fig04}(a)). This angle is eliminated between Eqs.~(\ref{ridge-ener-temp}) and (\ref{ridge-length}) to obtain the elastic energy as a function of the width $W$ and the length $\lambda$ of the ridge as assumed to derive Eqs.~(\ref{main-eq}):
\begin{equation}
\label{elast-ener}
U_{\rm{E}}(\lambda,W) = C_{\lambda}^7C_{{\rm R}}\, B t^2 W^5 \lambda^{-7}.
\end{equation}
Therefore we have
\begin{subequations}
\label{forces}
\begin{align}
\label{dlambda}
 F = -\partial_{\lambda} U_{\rm{E}} &= 7 C_{\lambda}^7C_{{\rm R}}\, B t^2 W^5 \lambda^{-8}, \\
\label{dw}
 \partial_W U_{\rm{E}} &= 5 C_{\lambda}^7C_{{\rm R}}\, B t^2 W^4 \lambda^{-7}.
\end{align}
\end{subequations}

The remaining quantity to determine before computing the crack paths is the angle $\varphi$. From Fig.~\ref{fig01}(e), it is expected that $b/W \ll 1$ leading to $\sin \varphi \simeq 4b/W$ and $\cos \varphi \simeq 1$. When $W$ decreases as the two cracks get closer, the ratio $b/W$ could, a priori, increases to reach values of order 1. However, we show in the Appendix that $b\simeq \lambda \alpha/4$ which combined with Eq.~(\ref{ridge-length}) gives 
\begin{equation}
\label{cond-b}
b/W \simeq C_{\lambda}^3 W t/(4\lambda^2).
\end{equation}
Therefore, we have to evaluate this quantity a posteriori, once $\lambda$ is known, to verify that it is indeed small. We assume $b/W \ll 1$ for the moment and we verify below the consistency of this assumption. We thus have
\begin{equation}
\label{phi}
\sin \varphi \simeq C_{\lambda}^3 \, W t \lambda^{-2} \quad {\rm and} \quad \cos \varphi \simeq 1. 
\end{equation}

Using Eqs.~(\ref{force}) and (\ref{forces}) together with the expression of the angle $\varphi$ (\ref{phi}), we obtain the equation giving the length of the ridge
\begin{equation}
\label{eq-lam}
7C_{\lambda}^7C_{{\rm R}}\, B t^2 W^5 \lambda^{-8}+5 C_{\lambda}^{10} C_{{\rm R}}\, B t^3 W^5 \lambda^{-9}=\gamma t \cos \theta.
\end{equation}
Depending on which of the two terms of the left-hand side of Eq.~(\ref{eq-lam}) dominates, we get two different regimes. The first term dominates when 
\begin{equation}
\label{cond-lambda}
\lambda \gg (5C_{\lambda}^3/7)\, t
\end{equation}
which is expected to be the dominant regime. Physically, $\lambda$ cannot be smaller than the film thickness. Therefore, the second term never dominates but its influence increases as $\lambda$ approaches $t$ and can be estimated by neglecting the first term. Since the second term encodes the influence of the transverse curvature of the ridge, we see that it is essentially negligible.

\section{Scaling for $W \gg W_{{\rm c}}$}
\label{sec:main-scaling}

Neglecting the second term of the left-hand side of Eq.~(\ref{eq-lam}), we obtain
\begin{equation}
\label{lambda-1}
\lambda = \left[7C_{\lambda}^7C_{{\rm R}}\, B t W^5/(\gamma \cos \theta)\right]^{1/8}.
\end{equation}
Even if we have neglected the term containing $\sin \varphi$, we still have to consider the condition ensuring the validity of Eq.~(\ref{phi}), namely $b/W\ll 1$, because we have to verify the validity of the condition (\ref{cond-lambda}) involving both terms of Eq.~(\ref{eq-lam}). The condition $b/W\ll 1$ is verified explicitly by using Eqs.~(\ref{cond-b}) and (\ref{lambda-1}) and leads to the equivalent condition $W\gg W^{\star} \sim \gamma/E \sim 1$ $\mu$m\footnote{We use the symbol $\sim$ when prefactors of order 1 are dropped and the symbol $\simeq$ when higher order terms are neglected.} where we used some typical values for bidirectional polypropylene films employed in the experiments~\cite{hamm08,baya11}. The domain of validity of the approximation consisting in neglecting the second term of Eq.~(\ref{eq-lam}) is made explicit by using Eqs.~(\ref{cond-lambda}) and (\ref{lambda-1}):
\begin{equation}
\label{wc}
W \gg W_{{\rm c}} \sim t \left(\frac{\gamma}{Et}\right)^{1/5}.
\end{equation}
The length $W_{{\rm c}}$ is the distance between the two crack tips at which the exponent of the power law characterizing the crack paths changes from $8/11$ to $9/8$ as $W$ decreases, see below. 

The length $W^{\star}$ fixes the domain of validity of Eq.~(\ref{eq-lam}) which is derived by assuming $b/W \ll 1$. The length $W_{{\rm c}}$ fixes the domain of validity of the approximation used in this section where one term of Eq.~(\ref{eq-lam}) is neglected. Therefore, the condition $W \gg W^{\star}$ must always be satisfied to obtain consistent results. Since the regime discussed in this section is valid for $W \gg W_{{\rm c}}$, the condition $W \gg W^{\star}$ is certainly verified if $W^{\star} < W_{{\rm c}}$. This last inequality sets a limit on the film thickness: $t > t^{\star} \sim \gamma/E \sim 1$ $\mu$m. This limit is satisfied in the experiments we consider.

\begin{figure}
\includegraphics[width=\columnwidth]{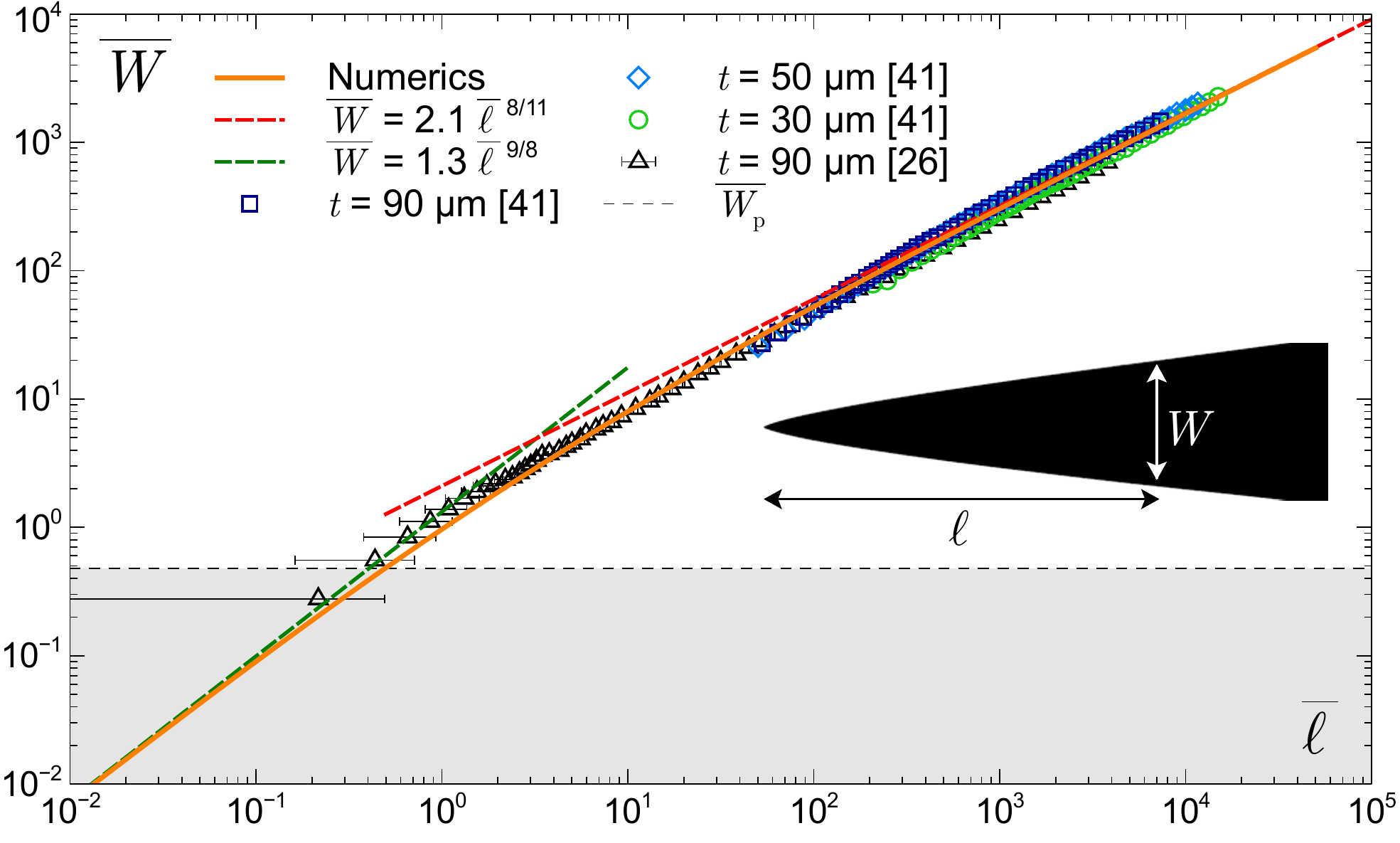}
\caption{(color online) Comparison between rescaled data, scalings (\ref{scalings}) and the numerical solution with $C_{\lambda}\simeq 1.45$. Data from Ref.~\cite{baya11} are rescaled using $E=2.2$ GPa, $K=2.6$ MPa m$^{1/2}$~\cite{baya11} and $\gamma=K^2/E$~\cite{freu98}. Data from Ref.~\cite{roma13} are rescaled using $B=1.5\, 10^{-4}$ Nm, $\gamma t = 1.9$ N~\cite{hamm08} and $\nu=0.3$~\cite{bros96}. The error on the position of the flap tip is 10 $\mu$m. The shaded region corresponds to $\overline{W}\lesssim \overline{W}_{{\rm p}}\simeq 0.48$. The inset shows a detached flap obtained once the two cracks meet.}
\label{fig02}
\end{figure}

The direction followed by the crack path is obtained by combining Eq.~(\ref{theta}) with Eqs.~(\ref{dw}), (\ref{phi}) and (\ref{lambda-1}) 
\begin{equation}
\label{theta-1}
\tan \theta [\cos \theta]^{\frac{1}{8}} = 5 \left[\frac{C_{\lambda}}{7}\right]^{\frac{7}{8}} \left[\frac{C_{{\rm R}}}{12(1-\nu^2)} \frac{Et}{\gamma}\right]^{\frac{1}{8}}  \left[\frac{t}{W}\right]^{\frac{3}{8}}.
\end{equation}
This equation shows that for large $W$, as considered in this section, $\theta$ is small and the left-hand side can be approximated by $\tan \theta$. We can now solve Eq.~(\ref{ode}) to obtain the crack path equation
\begin{equation}
\label{path-1}
\frac{W(\ell)}{t} = 1.56 \left[\frac{C_{\lambda}^7 C_{{\rm R}}}{(1-\nu^2)}\right]^{\frac{1}{11}} \left[\frac{Et}{\gamma}\right]^{\frac{1}{11}} \left[\frac{\ell}{t}\right]^{\frac{8}{11}}.
\end{equation}
The exponent $8/11 \simeq 0.73$ is very close to the exponent measured experimentally and fits quite well the data for large $W$, see Fig.~\ref{fig02}. The prefactor $Et/\gamma$ depends on the material constants and reflects the competition between elastic and fracture energies as expected. The fracture energy favors straight crack paths with $\theta=0$ to minimize the crack length whereas the elastic energy favors $\theta = \pi/2$ in order to reduce the width of the ridge as ``quickly" as possible. Equation (\ref{theta-1}) shows these tendencies with $\theta \to 0$ as $\gamma \to \infty$ and $\theta \to \pi/2$ when $Et \to \infty$. The small value of the prefactor exponent ($1/11$) explains why a simple rescaling by the film thickness leads nevertheless to a good collapse of the data~\cite{baya11}.

\section{Scaling for $W \ll W_{{\rm c}}$}
\label{sec:sec-regime}

As mentioned above, this regime is never fully reached since the length of the ridge cannot be smaller than the film thickness. The exponent derived here may thus be viewed as an asymptotic limit. The crack path exponent near the tip of the detached flap should approach this limit. This regime is described by neglecting the first term of the left-hand side of Eq.~(\ref{eq-lam}) which gives
\begin{equation}
\label{lambda-2}
\lambda = \left[5C_{\lambda}^{10} C_{{\rm R}}\, B t^2 W^5/(\gamma \cos \theta)\right]^{1/9}.
\end{equation}
The condition $b/W\ll 1$ is verified by using Eqs.~(\ref{cond-b}) and (\ref{lambda-2}) and leads to the equivalent condition $W\gg W^{\ast} \sim t [\gamma/(Et)]^2 \sim 10^{-2} \mu$m for  $t\sim 50$ $\mu$m. Smaller values of $W$ are described by the regime $b/W\gg 1$. However, the spatial resolution of the experiments being typically limited to 10 $\mu$m~\cite{roma13,baya11}, we do not discuss this marginal regime. The domain of validity of the approximation consisting in neglecting the first term of Eq.~(\ref{eq-lam}) is obtained by using Eq.~(\ref{cond-lambda}) with the reverse inequality sign and Eq.~(\ref{lambda-2}). We obtain $W \ll W_{{\rm c}}$ where $W_{{\rm c}}$ is exactly the same, prefactor included, than the one obtained in Eq.~(\ref{wc}) as it should. The necessary condition $W^{\ast} < W_{{\rm c}}$ sets the same limit on the film thickness than in the previous regime: $t > t^{\star}$.

The direction followed by the crack path is obtained by combining Eq.~(\ref{theta}) with Eqs.~(\ref{dw}), (\ref{phi}) and (\ref{lambda-2})
\begin{equation}
\label{theta-2}
\tan \theta [\cos \theta]^{\frac{2}{9}} = C_{\lambda}^{-\frac{7}{9}} \left[\frac{5C_{{\rm R}}}{12(1-\nu^2)} \frac{Et}{\gamma}\right]^{\frac{2}{9}}  \left[\frac{W}{t}\right]^{\frac{1}{9}}.
\end{equation}
This equation shows that for small $W$, as considered in this section, $\theta$ is small and the left-hand side can again be approximated by $\tan \theta$. The crack path equation is obtained by solving Eq.~(\ref{ode}) using Eq.~(\ref{theta-2}):
\begin{equation}
\label{path-2}
\frac{W(\ell)}{t} = 1.53 \left[\frac{C_{\lambda}^{-\frac{7}{2}} C_{{\rm R}}}{(1-\nu^2)}\right]^{\frac{1}{4}} \left[\frac{Et}{\gamma}\right]^{\frac{1}{4}} \left[\frac{\ell}{t}\right]^{\frac{9}{8}}.
\end{equation}
The exponent increases from $8/11$ to $9/8$ as the distance $W$ between the two crack tips tends to zero in reasonable agreement with data, see Fig.~\ref{fig02}. Notice that this regime is difficult to probe experimentally because it is close to the experimental spatial resolution~\cite{roma13,baya11}. 

\section{Comparison with data}
\label{sec:data}

Equations~(\ref{path-1}) and (\ref{path-2}) giving the crack paths in the two identified regimes are rescaled as follows
\begin{equation}
\label{rescalings}
\overline{W} = \left[\frac{Et}{\gamma} \right]^{\frac{1}{5}} \frac{W}{t} \quad {\rm and} \quad \bar{\ell} = \left[\frac{Et}{\gamma} \right]^{\frac{2}{5}} \frac{\ell}{t},
\end{equation}  
to obtain
\begin{subequations}
\label{scalings}
\begin{align}
\label{large-w}
\overline{W} &= C_1\, \bar{\ell}^{\, \frac{8}{11}} \quad {\rm for}\quad \overline{W} \gg \overline{W}_{{\rm c}}, \\
\label{small-w}
\overline{W} &= C_2\, \bar{\ell}^{\, \frac{9}{8}} \quad {\rm for}\quad \overline{W} \ll \overline{W}_{{\rm c}},
\end{align}
\end{subequations}
where
\begin{align}
\label{k-i}
& C_1=1.56 \left[\frac{C_{\lambda}^7 C_{{\rm R}}}{(1-\nu^2)}\right]^{\frac{1}{11}},\quad C_2=1.53 \left[\frac{C_{\lambda}^{-\frac{7}{2}} C_{{\rm R}}}{(1-\nu^2)}\right]^{\frac{1}{4}}, \\
& \overline{W}_{{\rm c}}=0.65 \left[(1-\nu^2)C_{\lambda}^{17}C_{{\rm R}}^{-1}\right]^{\frac{1}{5}}, \quad C_{{\rm R}} = {\cal R}[12(1-\nu^2)]^{\frac{1}{6}}, \nonumber
\end{align}
where ${\cal R} = 1.20\pm 0.04$~\cite{lobk96}. Figure~\ref{fig02} shows a nice collapse of the data rescaled with Eqs.~(\ref{rescalings}) together with a good agreement with Eqs.~(\ref{scalings}). The value $C_1=2.1$ is obtained from a fit of the data for large $\bar{\ell}$ which implies, from Eq.~(\ref{k-i}), $C_{\lambda} \simeq 1.45$ for $\nu=0.3$~\cite{bros96}. As expected, the free parameter is of order 1. The parameter $C_2\simeq 1.3$ is then computed from Eq.~(\ref{k-i}).

In order to obtain the evolution of $\overline{W}$ when it is of order~1, we need to solve the problem numerically because both terms of the left-hand side of Eq.~({\ref{eq-lam}) have the same order of magnitude. For this purpose, we rescale $W$ using Eq.~(\ref{rescalings}) and define $\bar{\lambda} = \lambda/t$. Equation~(\ref{eq-lam}) becomes
\begin{equation}
\label{alg-const}
7 A\, \overline{W}^{\, 5} \bar{\lambda}^{-8} + 5 A C_{\lambda}^3\, \overline{W}^{\, 5} \bar{\lambda}^{-9}=1,
\end{equation}
where $A=C_{\lambda}^7 C_{{\rm R}}/(12(1-\nu^2))$ and where we set $\cos \theta =1$ since $\theta$ is always small as seen above. Equation (\ref{theta}) together with Eqs.~(\ref{ode}), (\ref{dw}), (\ref{phi}) gives the following differential equation
\begin{equation}
\label{ode-rescaled}
{\rm d} \overline{W}/{\rm d} \bar{\ell}= 10A \, \overline{W}^{\, 4} \bar{\lambda}^{-7},
\end{equation}
where $\ell$ has been rescaled using Eq.~(\ref{rescalings}) and $\sin \theta$ has been replaced by $\tan \theta$. The differential equation (\ref{ode-rescaled}) is thus supplemented by an algebraic constraint (\ref{alg-const}). This semi-explicit differential-algebraic equation is easily solved numerically using, for example, Mathematica. The resulting crack path is reported in Fig.~\ref{fig02} and describes well the data.

The quantities $W^{\star}$, $W^{\ast}$ and $t^{\star}$ set limits for the mathematical consistency of the model; they are all satisfied for the experiments we consider. The quantity $W_{{\rm c}}$ is a limit separating the two identified regimes; one of them being only asymptotic. Physically, the length of the ridge is expected to be limited by the film thickness, namely $\bar{\lambda} \gtrsim 1$. Equation (\ref{alg-const}) imposes then $\overline{W}\gtrsim \overline{W}_{{\rm p}} = [A(7+5C_{\lambda}^3)]^{-1/5}$ which provides a physical limit of this model, see Fig.~\ref{fig02}.

\section{Smaller peeling angles}
\label{sec:small}

Finally, we discuss briefly the situation where the peeling angle is smaller than $\pi$. We expect that a Lobkovsky-Witten ridge emerges only for a peeling angle close to $\pi$ such that the fold joining the cracks possesses a transverse curvature and pinched edges. For smaller peeling angles $\phi$, the two crack tips should no longer be points of high curvature in the sheet (compared to the average curvature in the ridge). Therefore, the ridge joining them should be similar to the one occurring for adhesive sheets~\cite{hamm08,krug11} and should contain only bending energy. The energy of such a ridge has been computed in Ref.~\cite{roma13}:
\begin{equation}
\label{elast-ener-ben}
U_{\rm{E}}(\lambda,W) = f(\phi) \frac{B W}{\lambda}, \quad f(\phi)=4 [1-\cos(\phi/2)]^2,
\end{equation}
where $B$ is the bending modulus of the sheet, $\lambda$ and $W$ are the length and the width of the ridge respectively. Therefore, instead of Eq.~(\ref{forces}), we have now 
\begin{equation}
\label{force-ben}
F = -\partial_{\lambda} U_{\rm{E}} = f(\phi) \frac{B W}{\lambda^2} \quad \text{and} \quad \partial_W U_{\rm{E}} = f(\phi) \frac{B}{\lambda}.
\end{equation}
For such a ridge, there is no transverse curvature ($\varphi=0$). Equation (\ref{force}) together with Eq.~(\ref{force-ben}) fixes the ridge length as
\begin{equation}
\label{lambda-ben}
\lambda=\left[f(\phi)B W/(\gamma t \cos \theta)\right]^{1/2}.
\end{equation}
The crack path direction is obtained by combining Eq.~(\ref{theta}) with Eq.~(\ref{force-ben}) and using Eq.~(\ref{lambda-ben})
\begin{equation}
\tan \theta [\cos \theta]^{\frac{1}{2}}=\left[f(\phi)B/(\gamma t W)\right]^{1/2}.
\end{equation}
This equation shows that for large $W$ ($W \gtrsim 1$ mm), $\theta$ is small and the left-hand side can be approximated by $\tan \theta$. We can now solve Eq.~(\ref{ode}) to obtain the crack path equation
\begin{equation}
\label{path-ben}
\frac{W(\ell)}{t}=\left[\frac{3f(\phi)}{4(1-\nu^2)}\right]^{\frac{1}{3}} \left[\left(\frac{Et}{\gamma}\right)^{\frac{1}{2}} \frac{\ell}{t} \right]^{\frac{2}{3}}.
\end{equation}
We notice that the peeling angle does not affect the exponent of this scaling as for the tearing of adhesive sheets~\cite{roma13,krug11}. Experiments for various peeling angles smaller than $\pi$ are needed to test this scaling. 

\begin{figure}
\includegraphics[width=\columnwidth]{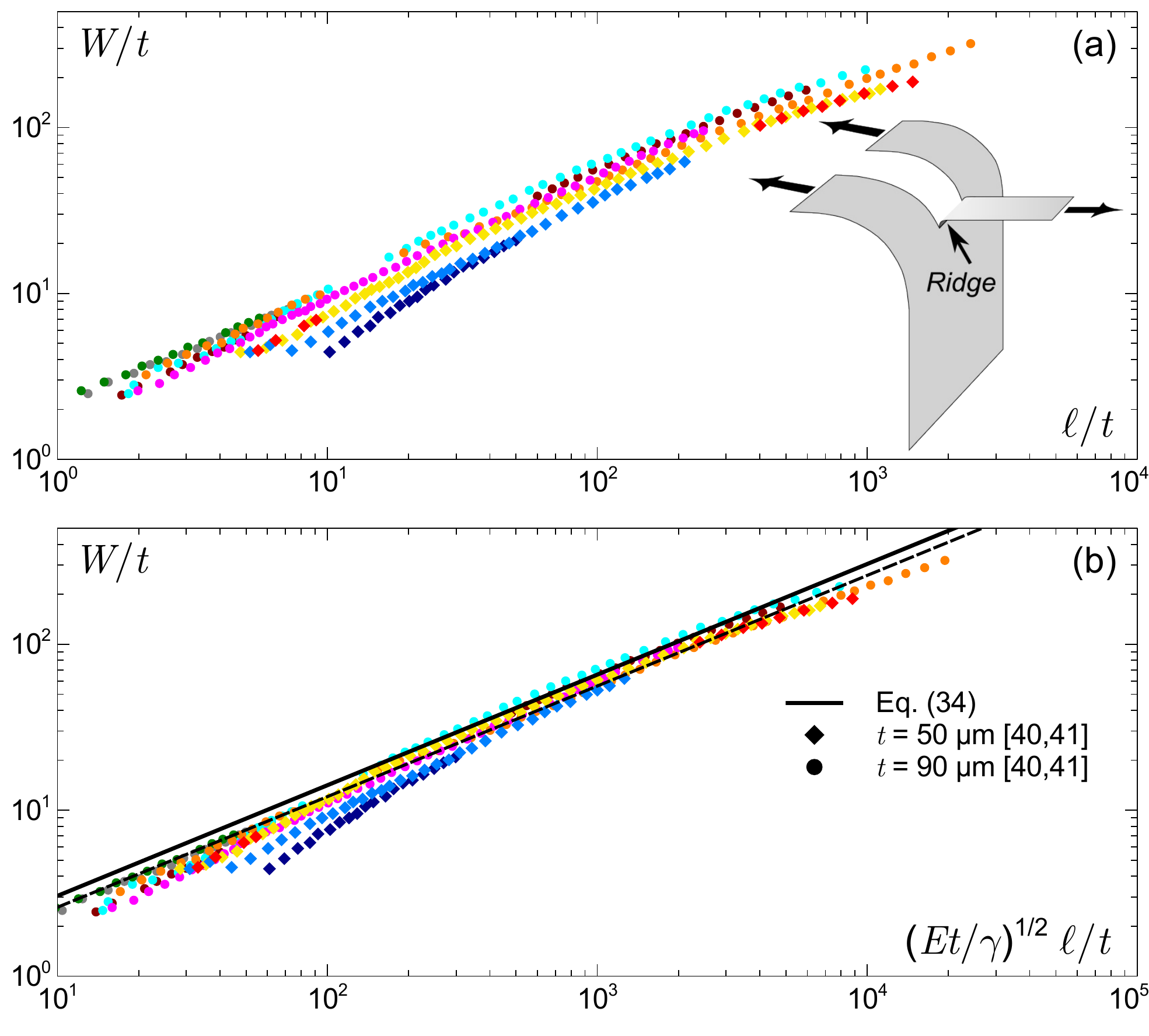}
\caption{(color online) (a) Data for the crack paths in the trousers configuration from Refs.~\cite{baya10,baya11} rescaled by the sheet thickness. Inset: Schematic of the trousers configuration where large arrows indicate the directions of the applied forces. (b) Same data rescaled by the scaling obtained in Eq.~(\ref{path-ben}). Data are rescaled using $E=2.2$ GPa, $K=2.6$ MPa m$^{1/2}$~\cite{baya11} and $\gamma=K^2/E$~\cite{freu98}. The crack path obtained from Eq.~(\ref{path-ben}) is shown for $\phi=90^{\circ}$ (solid line) and $\phi = 80^{\circ}$ (dashed line). (a)-(b) The different colors correspond to several realization of the same experiment.}
\label{fig03}
\end{figure}

Notice however that this scaling with $\phi=\pi/2$ should describe well the data obtained in Refs.~\cite{baya10,baya11} in the ``trousers" configuration, see inset of Fig.~\ref{fig03}(a). Indeed, in this configuration, we expect that the crack tips should not be points of high curvature in the sheet and the ridge joining them should also be described by Eq.~(\ref{elast-ener-ben}). The bending energy stored in the sheet outside this ridge should not affect significantly the crack path. Figure~\ref{fig03} shows the evolution of the distance between the two crack tips as a function of the distance from the point where they meet. In Fig.~\ref{fig03}(a), the data are rescaled by the sheet thickness, as proposed in Ref.~\cite{baya11}, whereas in Fig.~\ref{fig03}(b), the data are rescaled according to the scaling obtained in Eq.~(\ref{path-ben}). A better collapse of the data is obtained in Fig.~\ref{fig03}(b) compared to Fig.~\ref{fig03}(a) together with a good agreement with Eq.~(\ref{path-ben}) without any fitting parameter.

\section{Conclusions}

We have shown how the energies focused in the tip of the cracks and in the elastic ridge joining them act together in a non trivial way to produce characteristic crack paths described by a power law with an exponent $8/11$ and a prefactor reflecting the competition between elastic and fracture energies, see Eq.~(\ref{path-1}). The close agreement with experiments is shown in Fig.~\ref{fig02}. In addition, a second regime, induced by the transverse curvature of the ridge, occurs for small distances between the crack tips. This regime is only asymptotic but slightly modifies the crack path such that the exponent of the power law increases to reach values close to $9/8$, see Eq.~(\ref{path-2}) and Fig.~\ref{fig02}. A global rescaling has been found and leads to Eqs.~(\ref{rescalings})-(\ref{k-i}). The governing equation has also been solved numerically to obtain the complete crack path, beside its asymptotic scalings, with a good agreement with experiments using only one free parameter of order 1. 

\appendix

\section{Lobkovsky-Witten ridge}

We consider a sheet of width $W$ in the configuration shown in Fig.~\ref{fig04} with a fixed dihedral angle $\alpha$. We consider the limit of small dihedral angle where the curved parts of the ridge shown in Fig.~\ref{fig04}(a),(c) can be approximated by arc of circles. Notice that the scalings (\ref{ridge}), obtained from a boundary layer analysis, are not restricted to small values of $\alpha$~\cite{lobk96}. 

From the triangles $ABD$, $BCD$ and $ABC$ of Fig.~\ref{fig04}(b), we have respectively $\cos \alpha = R_1/(R_1+b)$, $\cos \alpha = (R_1-a)/R_1$ and $\tan \alpha \simeq \alpha = 2(b+a)/h$. The two first relations implies $a=b\cos \alpha \simeq b$ and $R_1=b \cos \alpha/(1-\cos \alpha) \simeq 2b/\alpha^2$. We thus obtain $\alpha \simeq 4b/h$. In addition, we have $\lambda = 2\alpha R_1$ which is equivalent to $b\simeq \lambda \alpha/4$ as mentioned in the main text. Using the expression of $\alpha$ obtained above, we also have $\lambda \simeq h$. 

The energy, $U_{\rm{E}}$, of the ridge is composed essentially of a bending energy, $U_{{\rm b}}$, in the longitudinal direction along its length $\lambda$ ($z$-axis) and a stretching energy, $U_{{\rm s}}$, in the transverse direction along its width $W$ ($y$-axis), see Fig.~\ref{fig04}(a). These energies are localized in a region of area $S \sim \lambda W$. From Fig.~\ref{fig04}(b), the longitudinal curvature is given by $\kappa \sim a/h^2 \sim b/\lambda^2$. Therefore, the bending energy reads
\begin{equation}
U_{{\rm b}} \sim E t^3 (b/\lambda^2)^2 S \sim E t^3 \alpha^2 W/\lambda.
\end{equation}
The stretching is due to the sag of the ridge inducing an increase in length along its width of order $(b/W)^2$. The stretching energy thus reads
\begin{equation}
U_{{\rm s}} \sim E t (b/W)^4 S \sim E t\, \alpha^4 \lambda^5/W^3.
\end{equation}
Upon minimization of the total energy $U_{\rm{E}}= U_{{\rm b}}+U_{{\rm s}}$ with respect to $\lambda$ ($\partial U_{\rm{E}}/\partial \lambda =0$), we obtain the scalings~(\ref{ridge}) of the main text.

\begin{figure}
\includegraphics[width=0.9\columnwidth]{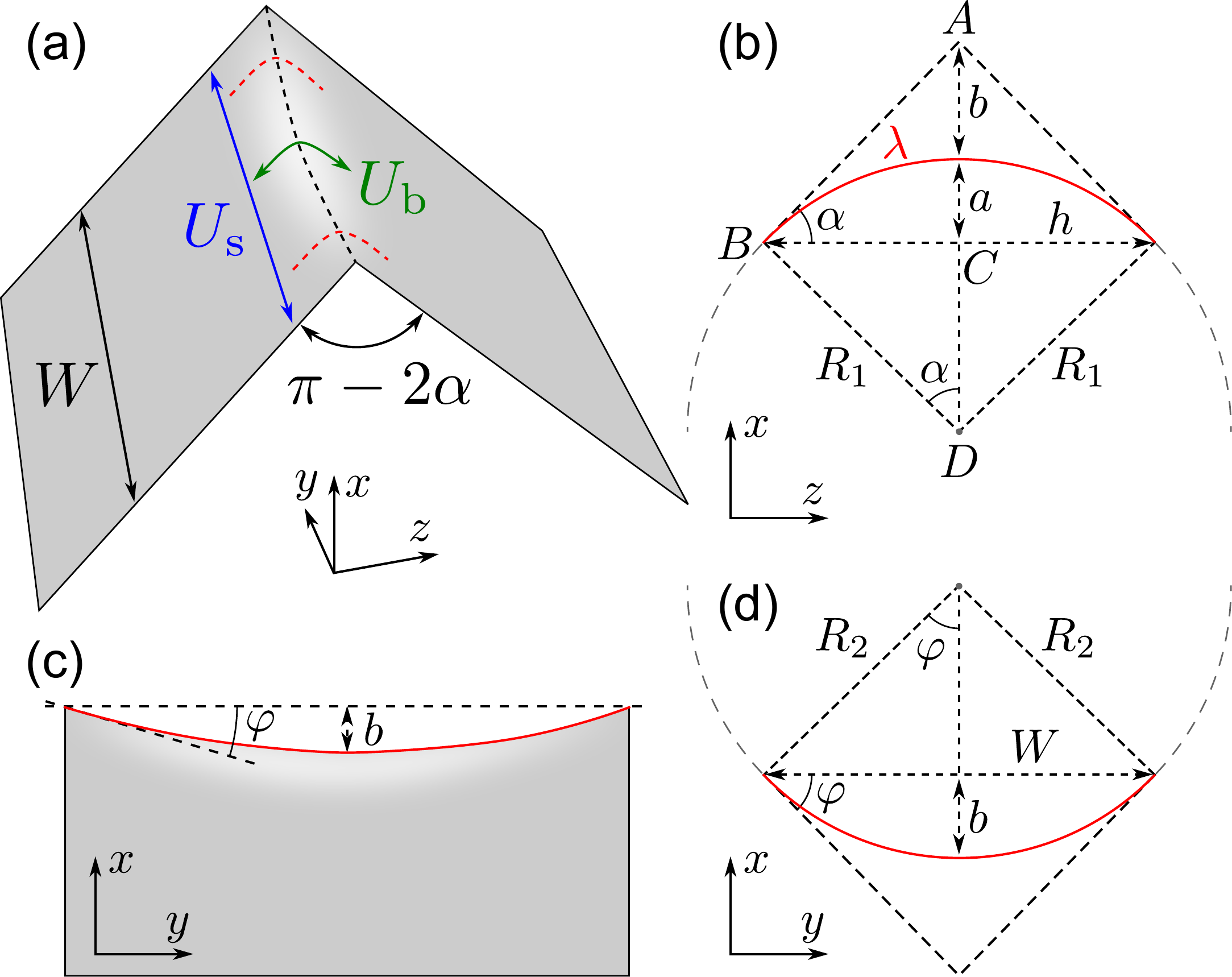}
\caption{(color online) Schematics of the Lobkovsky-Witten ridge.}
\label{fig04}
\end{figure}

The angle $\varphi$ originating from the sag of the ridge is computed from Fig.~\ref{fig04}(d) where we have $R_2 = W^2/(8b)+b/2$ and $\sin \varphi = W/(2R_2) \simeq 4b/W$ at the first order in $b/W$. Using the expression of $b$ obtained above, we have $\sin \varphi\simeq \lambda \alpha /W$. Using the expression (\ref{ridge-length}) of $\alpha$, we obtain Eq.~(\ref{phi}).

\acknowledgments
We thank B. Roman for discussions about experiments and PRODEX for financial support.

\end{document}